\documentclass[12pt,letterpaper]{article}
\UseRawInputEncoding
\usepackage{amsmath,pgf,pgfarrows,pgfnodes,float,appendix, hyperref,scalerel,amssymb}
\usepackage{graphicx}
\usepackage{subfigure}

\usepackage[margin=0.9in]{geometry}
\usepackage{pgfplots}
\usepackage{amsmath}
\usepackage{tikz}

\title{{\rm\footnotesize \qquad \qquad \qquad \qquad \qquad \ \qquad \qquad \qquad \ \ \ \ \ \                   }\vskip.5in Fluctuating Hydrodynamics and Wormholes}
\author{\\ Thomas Banks
Department of Physics and NHETC\\
Rutgers University, Piscataway, NJ 08854\\
E-mail: \href{mailto:tibanks@ucsc.edu}{tibanks@ucsc.edu}}

\date{}
\begin{document}
\maketitle

\begin{abstract} 
We show that a recent reformulation of hydrodynamic equations for a large class of models consisting of q-dits on a graph with short range interactions is sufficient for understanding chaotic behavior.  Any such system consists of large subsystems coupled together by interactions whose relative strength goes to zero with the subsystem size.  In the absence of conservation laws other than energy, the Hamiltonians of the subsystems form a complete set of commuting operators. 
The hydrodynamic variables are the block diagonal matrix elements $\rho (e(X))$ of the density matrix in the joint eigenbasis of the subsystem Hamiltonians, averaged over energy bins.  To leading order in the inverse subsystem size, $\rho (e(X) , t)$ satisfies a classical stochastic equation, which for certain systems takes the form of a functional Fokker-Planck equation. In such systems the time averaged spectral form factors can be written as a two dimensional Euclidean functional integral, on a space with multiple  disconnected boundaries. The failure of factorization in this representation is attributable to the time averaging necessary to apply the hydrodynamic approximation. The bulk Euclidean action is purely topological. We make tentative explorations of the special properties of the system that are required in order to have a representation as a functional integral over metrics.  \end{abstract}

\section{Introduction}

There has been much recent work demonstrating that semi-classical Euclidean path integrals over metrics encode a lot of information about the level statistics of black hole systems, greatly expanding on the seminal work of Gibbons and Hawking, who calculated the entropy of black holes and de Sitter horizons in this manner.  One of the puzzles posed by this work is the lack of factorization of the relevant semi-classical expressions for quantities like the {\it spectral form factor} (SFF) $| {\rm Tr}\ e^{-(\beta + i t) H }|^2 $.  In \cite{lindil} I opined that these calculations should probably be understood in the context of Jacobson's\cite{ted95} derivation of Einstein's equations as the hydrodynamic equations of the Bekenstein-Hawking (BH) entropy law applied to arbitrary causal diamonds.   They are the local expression of the first law of thermodynamics.  In the present paper we will explore what the hydrodynamics of a large class of quantum systems tells us about the spectral form factor.   Much of what we have to say was anticipated in a beautiful paper by Swingle and Winer\cite{swingle}.  These authors used the effective field theory of hydrodynamics\cite{liurangamanietal}, whereas our considerations are based on the more microscopic derivation of hydrodynamics presented in\cite{tblucas}.   

The general class of systems we will study have Hamiltonians of the form
\begin{equation} H = \sum_X H(X) + \sum_{X,Y} H(X,Y)  . \end{equation}  Here 
\begin{equation} [H(X), H(Y)] = 0 , \end{equation} and in the present paper we will assume that they are a complete set of commuting operators.  This is tantamount to saying that the system has no other locally conserved quantities besides energy\footnote{And possibly a finite dimensional space of global/topological quantum numbers that do not act locally on the operator algebra.}.  The labels $X,Y$ refer to subsystems, each of which contains a large finite number of independent q-bits\footnote{If we want to deal with Hilbert spaces whose dimension is not a power of $2$ we simply add terms to the Hamiltonian which constrain the system to subspaces.  For example, we can write a q-trit as a two fermion system with a term in the Hamiltonian that penalizes states with occupation number $2$.}  We assume the dimensions of the subsystem Hilbert spaces are roughly the same.  The size of a typical eigenvalue of $H(X)$ is a large number that we write as $L^d$, with $L\gg 1$.   The Hamiltonians $H(X,Y)$ have two properties.  First, the norm of $ H(X,Y)$ is no larger than $L^{d - 1}$.  Secondly, for each $X$, $H(X,Y)$ is non-zero for only a finite number of subsystems $Y$.  This gives the system the structure of a {\it graph}, with $X$ labelling the nodes of the graph and the allowed pairs $(X,Y)$ labelling its links.  

In\cite{tblucas} we emphasized the example of a lattice system with short range couplings.  The subsystems consisted of subregions containing $L^d$ points, where $L$ was much larger than the range of interactions.  The allowed pairs $(X,Y)$ were then links between nearest neighbors on a coarse grained lattice whose points were in the center of each subregion.  The restriction on the norm of $H(X,Y)$ follows from the geometry of the lattice.  It is a sum of local Hamiltonians over the surface separating $X$ from $Y$.  But the conclusions of\cite{tblucas} are much more general than lattice models.

The main conclusion of\cite{tblucas} was that the diagonal matrix elements of a generic density matrix in the basis where the local Hamiltonians $H(X)$ were diagonal, satisfied a Markov equation
\begin{equation} \frac{dP[e(X), t)}{dt} = - \int [de(Y)] W[e(X),e(Y)] P[e(Y),t)  , \end{equation} where we've treated the local averaged $H(X)$ eigenvalues, $e (X)$ as continuous\footnote{The mixed bracket indicates that $P [e(X), t) $ is a functional of the energy density and a function of $t$. }  . $e(X)$ is defined as the central value of the spectrum of $H(X)$ in an interval $[e(X) - \delta/2 , e(X) + \delta/2]$ .   The time scale $\delta t = \delta^{-1}$ is chosen to be shorter than the hydrodynamic time scale of the stochastic equation, but much longer than microscopic time scales of the system.  The Markov equation emerges from time averaging over the interval $\delta t$.

 $W$\cite{tblucas} is constructed from quantum transition probabilities between subspaces of fixed $e(X)$, over the time scale $\delta t$.  $W$ is of order $L^{-2}$.
This classical stochastic equation thus has a natural time scale $1/L^2$, where we've set the microscopic energy unit equal to $1$.  Hydrodynamic flows happen on a time scale much shorter than the inverse of the small eigenvalue differences of the Hamiltonian $H(X)$, which are of order $e^{-S(e(X))}$, with $S$ scaling like $L^d$.  It therefore makes sense to coarse grain the equation in energy, as we have done, over the scale $L^2 > \delta \gg D(X)^{-1}$, the dimension of the subsystem Hilbert space.  The local entropy density $S(e(X))$ is the logarithm of the dimension of the subspace with $e(X)  - \delta/2 < E(X) < e(X) + \delta/2 $.  In addition, over hydrodynamic time scales the energy $\sum_X e(X)$ is conserved.  $e(X)$ is effectively a continuous variable, whose value changes only via flow through the boundaries of $X$.  

Following Feynman\cite{feynhibbs} and Kac\cite{Kac} one can write a formal solution of this stochastic equation as a functional integral in imaginary time\footnote{The time variable is of course the real time of the system.  The word imaginary is useful because a Markov equation is mathematically identical to an imaginary time Schrodinger equation.  The resulting integrals are convergent.}.  The solution of the equation is
\begin{equation} \rho [e(X), t) = e^{- t W}  \rho [e(X), 0) . \end{equation}  $W$ is the Markov operator that computes the time derivative of $\rho$ as a linear functional of $\rho$. As usual, we break time up into small intervals $\Delta t$ of size much smaller than the hydrodynamic time scales of interest, but larger than $\delta^{-1}$.  We insert the resolution of the identity in the $| e (X) )$ basis between every pair of terms in the product of short time evolution operators.  The notation $| v )$ represents a vector in the linear space of functionals of $e (X)$. The result is
\begin{equation} \rho [e(X), t] = \int d[e(X,t)] e^{- \int_0^t ds L[e(X,s) ] } \rho [e(X), 0] .  \end{equation} Here
\begin{equation} L[e(X,s))  =  \Delta t W[e(X, s + \Delta t); e(X,s)]  . \end{equation}  The notation in this last 
equation is a bit tricky.  $L$ is a functional of two different spatial functions $e(X,s)$ 
and $e(X, s + \Delta t)$ , and a function of $s$.  In order to get something that looks like a "normal" functional integral, we have to assume that the short time propagation kernel $W$ has matrix elements only between energy densities that are close to each other.  That is $e(X, s + \Delta t)$ has a power series expansion in powers of $\Delta t$.  For models based on lattices, this follows from the fact that the Hamiltonians $H(X,Y)$ are sums of contributions from different points on the boundary, each of which acts on only a small number of degrees of freedom.  These terms do not make significant changes in the block averaged energies $e (X)$.  This turns the general Markov equation into a functional generalization of the Fokker-Planck equation.  

More directly, we can simply expand the Lagrangian $L$ in powers of $dE/dt$, obtaining
\begin{equation} L = V[e(X)] + \int [de(Y)]  A[ e(X)] \dot{e}(Y) + \int [de(Y)][de(Z)] G [e(Y), e(Z)] \dot{e}(Y)\dot{e}(Z) . \end{equation}   This is a generalization to infinite dimensional function space of a non-linear model.  $A$ is a one form in function space, and $G$ a metric.  We've left the variable $s$ implicit in this equation, since everything is invariant under time translations.  We can eliminate the $A$ term by imposing time reversal invariance, and this was done in\cite{tblucas}.  The entropy functional $S[E] = \sum_X S(e(X))$ is a universal contribution to the "potential" $V$.  Other terms in $V$ can come from the microscopic Hamiltonian. The entropic contribution arises because the "states" $ | e (X) ) $ in the Feynman-Kac derivation of the functional integral are really subspaces, with large dimension, of the microscopic Hilbert space.

We've thus written the solution for the hydrodynamic approximation to the time evolution of a generic initial density matrix as an imaginary time functional integral.  Recall that the hydrodynamic equations are based on three approximations.  The first is the existence of a complete basis of the Hilbert space in which the Hamiltonians of certain large subsystems are all diagonal, with energy exchanged only between each of these subsystems and a few "neighbors" via a Hamiltonian that is a small perturbation of the subsystem energy.  The second is that we average the density matrix over time scales $\sim \delta^{-1}$.    This implies that energy levels of $H(X)$ are grouped into large, almost degenerate subspaces of width $\pm \delta$ around each, effectively continuous value, $e(X)$, with dimension $e^{S(e(X))}$, thus defining the local entropy density.   Finally, we assume that over time interval $\delta t = \delta^{-1}$ transition probabilities between different values of $e(X)$ can be neglected unless those differences are very small.  In\cite{tblucas} the final assumption was derived from the fact that $H(X,Y)$ was a sum of small terms describing interactions across local portions of the boundary between $X$ and $Y$.

We've also assumed for simplicity that the system has no other continuous symmetries, so that the simultaneous eigenstates of $H(X)$ form a complete set of states.  It is straightforward to incorporate other symmetries.  Discrete symmetries and higher form symmetries modify hydrodynamics in ways that have not yet been worked out.  

\section{The Spectral Form Factor}

The spectral form factor 
\begin{equation} F(t) = |Z(\beta + i t)|^2 , \end{equation} where $Z(\beta)$ is the thermal partition function, is a simple diagnostic of quantum chaos in many body quantum systems.  After appropriate averaging, $F(t)$ is a smooth function of $t$ at fixed $\beta$, with three characteristic time scales\cite{sss}.  First, there is a rapid exponential decay down to a value smaller by a factor $\sim e^{-2S}$ , where $S$ is the entropy of the system\footnote{We'll see that in the models considered in this paper, it is more appropriate to think of $S$ as the entropy of one of the block subsystems $X$.}.  The decay time scale is a microscopic parameter.  This is followed by a slow linear rise, called {\it the ramp} to a value $e^{-S}$ smaller than the value $F(t = 0)$.  This takes a time of order $e^S$ in microscopic units.  After that, $F(t)$ remains roughly constant.  The time scale $e^S$ is the time required for the system to become sensitive to the typical small energy differences, $e^{-S}$ in microscopic units, characteristic of a generic Hamiltonian on a large Hilbert space.

In\cite{sss1} the ramp was calculated in terms of a "Euclidean Wormhole Solution" of the equations of motion for collective fields in the SYK model.  The meaning of the phrase in quotes is the following.   One employs the Schwinger-Keldysh path integral formalism to compute the absolute square of the partition function in terms of a product of path integrals.  These path integrals are semi-classical in terms of the collective variables $\Sigma_{ab} (t,s) , G_{ab}(t,s)$, where $\Sigma$ is a Lagrange multiplier enforcing the constraint $G_{ab} (t,s) = \frac{1}{n} \sum_i \psi_i^a (t) \psi_i^b (s)$.  In the S-K formalism there are two kinds of fermions, describing the two factors of the partition function in $F(t)$, and $a,b$ run over these two kinds.  The off diagonal collective fields
$G_{12}$ describe correlations between the two kinds, which are of course absent if we work with a fixed Hamiltonian and compute microscopic correlators.  However, ensemble averaging over Hamiltonians, or time averaging, can produce such correlations.  The wormhole solution is a non-zero semi-classical configuration for $G_{12}$ and so is appropriate for describing such averaged situations.  Indeed\cite{sss} shows that the smooth ramp only emerges after averaging over time or random couplings.

In\cite{tblucas} and the present paper, we study a fixed Hamiltonian.  However, hydrodynamic equations emerge only after time averaging.  We begin from an exact formula for the SFF\cite{verlinde}
\begin{equation} F(t_1 - t_2) = Z^2 (\beta) {\rm Tr}\ [\rho_{TFD} (t_1) \rho_{TFD} (t_2)] , \end{equation} where 
$\rho_{TFD} (t)$ is the pure density matrix on the tensor product of two system Hilbert spaces, corresponding to the quantum state
\begin{equation} | TFD , t \rangle = \sum_i e^{- \beta/2 E_i - i E_i t } |E_i\rangle_R  |\bar{E}_i\rangle_L . \end{equation} Here $| \bar{E} \rangle = T | E \rangle,$ where $T$ is an anti-unitary operator, which commutes with the system Hamiltonian $\sum_X H(X) + \sum_{X,Y} H(X,Y) $, and $t_{12} = t_1 - t_2$.   This expression for $F(t) $ is invariant under simultaneous shifts of $t_1$ and $t_2$.  We get the average value of $F(t)$ by averaging this expression over both $t_1$ and $t_2$ in an interval of width $\delta t$ around any time $t$.  As in\cite{tblucas} we choose $\delta t$ to be just a bit smaller (parametrically in L) than the hydrodynamic time scale $L^{-2}$. To leading order in $L^{-1}$, $\bar{\rho}_{TFD} (0)$ has matrix elements only between eigenspaces with the same values of the smeared block energies  $e(X)$(with energy smearing scale $(\delta t)^{-1}$. ) .   Thus, we can use the results of\cite{tblucas} to write Markov equations for probabilities of the TFD system to have local energies $e(X)$ .  Indeed, the two tensor factors in the TFD Hilbert space evolve independently under the microscopic Hamiltonian $H$, so the entanglement between the two factors remains the same.  

The TFD density matrix is 
\begin{equation} \rho_{TFD} (t) = \sum_{i} | E_i , \bar{ E}_i \rangle\langle  E_i , \bar{E}_i | e^{-\beta E + 2 i t} . \end{equation} This is a state in the tensor product of the Hilbert space of the system with itself, and $| \bar{E}_i \rangle = T | E_i \rangle$, where $T$ is an anti-unitary operator that commutes with the Hamiltonian.  
At $t = 0$ and to leading order in $1/L$ the density matrix $\rho_{TFD} (0)$ is approximately block diagonal in subspaces with different values of $e(X)$, so the arguments of\cite{tblucas} apply.  The time derivative of the block diagonal matrix elements is of order $1/L$ and generates off block diagonal matrix elements only, to leading order in $1/L$.  Thus, the change in the block diagonal matrix elements is of order $L^{-2}$ and has the form of a classical master equation relating the time derivative to transition probabilities from one block to another.   Energy differences smaller than $(\delta t)^{-1}$ are negligible over the hydrodynamic time scale $L^{-2}$.  Most of the energy differences in each block are of order $e^{- L^p}$ for some positive integer $p$.  The time variation of the probability to have a fixed set of values $e(X) $ satisfies a closed Markov equation.  The Markov transition function will have a series expansion in inverse powers of $L^{-1}$ and this equation describes the evolution of the probabilities $P[e (X)]$ for times shorter than $e^{S(e (X))} $, the microscopic time scale multiplied by the dimensions of the local subspace.    Time evolution can be viewed as evolution of either entropy density or energy density, once we know the local temperature  
\begin{equation} d e (X) = T( e (X) ) d S ( e (X) ) , \end{equation} which is determined by the microscopic spectrum.

  Thus, for time evolution over hydrodynamic time scales we can write
\begin{equation} \partial_t \bar{\rho}_{TFD} [e(X), \bar{e}(X), t) =\int [d{\cal }^{\prime} (X)] W[e(X), e^{\prime} (X)] W[\bar{e} (X), \bar{e^{\prime}} (X) ] \bar{\rho}_{TFD} [e^{\prime} (X), \bar{e^{\prime} }(X), t) .  \end{equation}   Note that the microscopic expression for the Markov kernel $W$ in\cite{tblucas} is constructed from transition probabilities and so is invariant under the time reversal operation $T$.  

We can now perform the usual Feynman-Kac solution of the Markov equation by an imaginary time functional integral by writing the operator $e^{ - W t}$ as a product over infinitestimal time intervals, inserting complete sets of intermediate states.  As a consequence of the entanglement structure of the TFD state we can write expectation values in the time averaged $\bar{\rho}_{TFD} (t_1) $ in terms of functional integrals over a single energy variable $e(X,t)$, where $t$ ranges over the complex Schwinger-Keldysh contour.    To compute the time averaged spectral form factor we have to compute the expectation value of $\bar{\rho}_{TFD} (t_2)$ .  Since the time averaged density matrices are no longer pure, this will give rise to a non-factorizing connected part to the SFF.  Both density matrices, acting on the space of linear functionals of $e(X), \bar{e} (X)$ are diagonal in the $e(X)$ basis, which means that the initial conditions for $e(X,t_1)$ and $e(X,t_2)$ in the Schwinger-Keldysh path integral solution, are identical.  This means that the two path integrals are glued together into the single "Janus-Pacman" contour of Fig. 1 .  

\begin{figure}[btp]
\begin{center}
\includegraphics[scale=0.45]{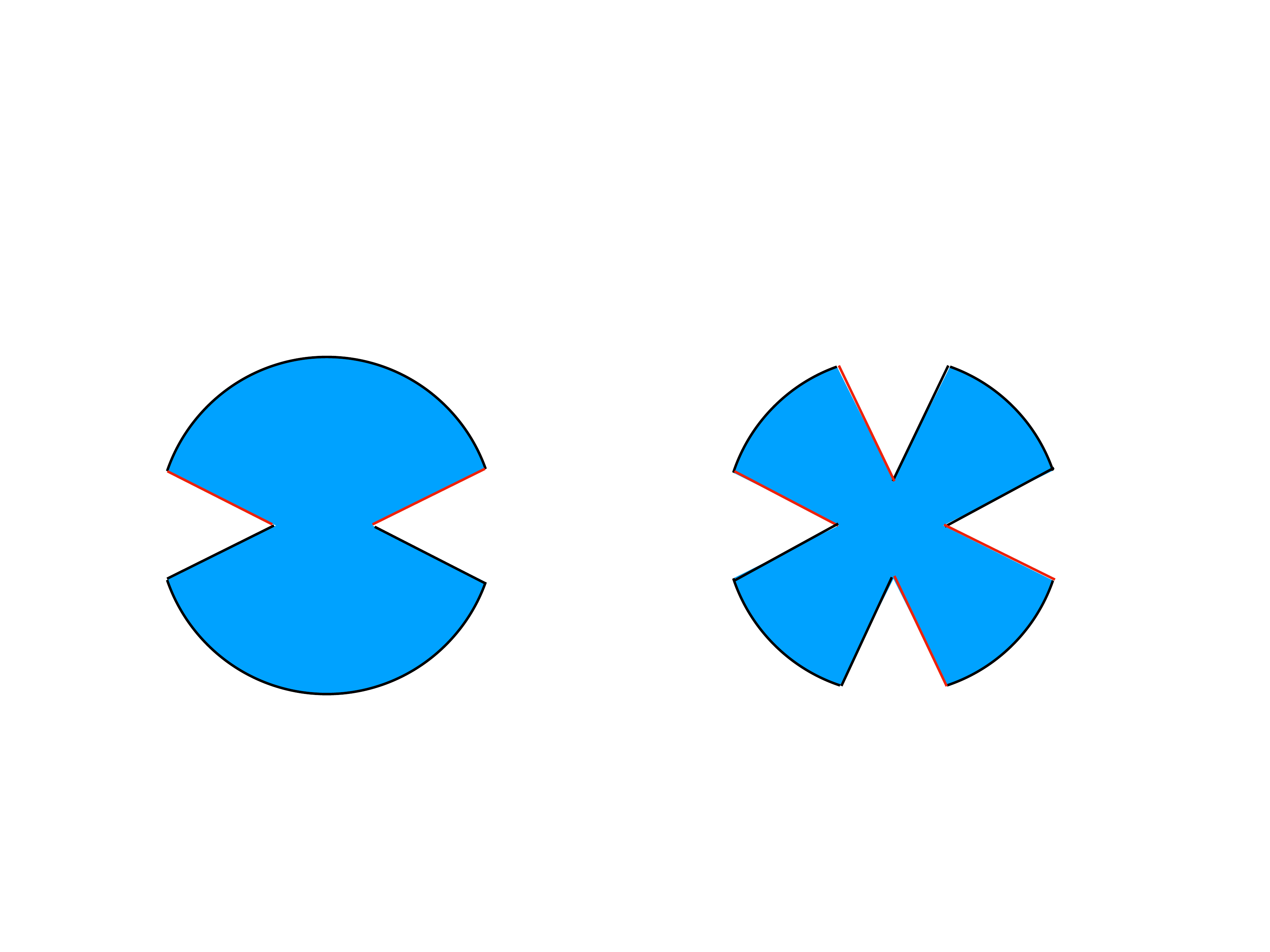}
\label{fig:1}
\vspace{-0.5cm}
\caption{The Contours for the Connected Parts of the Two and Four Point Spectral Correlation Functions. The blue regions are the two dimensional extension of those contours.}
\end{center}
\end{figure}

 Similarly, higher order correlation functions \begin{equation} {\rm Tr}\ [\bar{\rho}_{TFD} (t_1) \ldots \bar{\rho}_{TFD} (t_n)] \end{equation} will have a completely connected part given in terms of a functional integral expression over the multi-mouthed Pacman contour in the second half of that figure.  The reason for these violations of factorization is that the time averaged TFD density matrices are no longer pure\cite{verlinde}.  The action for the path integrals is independent of the number of mouths because it comes from solution of the a single stochastic equation, with thermal initial conditions.  This independence was the key assumption in the work of Swingle and Winer\cite{swingle}.  The action contains a universal term $- \sum_X S(e (X,t))$ and other terms with up to two time derivatives, which must be computed from the microscopic dynamics.  The microscopic Hamiltonian also determines the local temperature defined by the relation \begin{equation} \dot{e}(X,t) = T(e(X,t)) \dot{S} (X,t) . \end{equation}
This relation allows us to change variables in the functional integral from $e(X,t)$ to $S(X,t)$ .

The upshot of all of this is that, within the hydrodynamic approximation we can rewrite the spectral form factor in terms of path integrals over a variable $S(X,t)$ which describes the flow of entropy from large subsystem to large subsystem in the model.
The action for $S$ is a conventional quadratic kinetic term + potential action, to leading order in an expansion in $L^{-1}$.  In\cite{tblucas} the authors outlined ambiguities in defining higher order terms, but it is clear that, given some set of choices, one can continue this procedure to all orders in an expansion in $L^{-1}$ treating higher time derivative terms as corrections to an effective field theory: they should always be small perturbations and any situation in which they become large is one in which the expansion has broken down.  The connected parts of spectral form factors are exponentially small in this expansion, proportional to powers of $e^{- S}$.  

\subsection{From Pacman contours to two dimensional surfaces}

  The connected part of the two point SFF can be written as a product of two Euclidean functional integrals, while the common initial condition for the two $t_i$ values is a "wormhole" connecting them.  This connection occurs because the individual density matrices are diagonal in the space of linear functionals of $e(X)$, with identical initial conditions, and the exact formula for the SFF tells us to multiply them.  Verlinde\cite{verlinde} has shown how to write such path integrals in terms of actual two dimensional geometries.  If the effective action for $S(X,t)$ involves only first derivatives of $S$, then we can rewrite the Lagrangian as $L = P_S \dot{S} -  H(P_S, S) $, since the $P_S$ functional integral is ultra-local.  The first term can then be written (in many ways) as the integral of an exact two form over a two dimensional space-time, whose boundary is the two $t_i$ contours.   For the situation at hand, its topology is that of a cylinder connecting the two boundaries. The space-time has Euclidean signature, because there are no imaginary terms in the action.    The basic equations are 
\begin{equation} \int dt_1\  [P_S (t_1) \dot{S}(t_1)  +  \int dt_2\ P_S (t_2) \dot{S}(t_2)] = \int_{M_2}d( P_S (z) S e(z) ) , \end{equation} where $P_S (z)$ and $S(z)$ are smooth functions on $M_2$ whose boundary values on the two boundaries coincide with the original dynamical variables.  The common initial condition of the two Feynman-Kac path integrals can be taken to be the values of $e(z)$ on any one cycle homologous to the two boundaries, on the topological cylinder $M_2$.  The Hamiltonian terms in the original action, in which the detailed dynamics of particular systems resides, are pure boundary terms. Among these there is a universal factor $e^{S}$.  The bulk action is thus a topological field theory, of a particularly simple kind.  Note that in the last few equations, we've left the subsystem label $X$ and the sums over it in the action, implicit.  We can repeat this argument for the multi-boundary Pacman contours, obtaining the same topological field theory on a surface with higher genus topology and no handles.

Topological field theory is independent of the geometry on the two dimensional space defined by Verlinde's trick, but we can choose a particular geometry by choosing the continuation of the $S$ field into the bulk to have an imaginary part $S_I$ that vanishes on the boundary and appears in the two dimensional action in the form
\begin{equation} \Delta I = i \int_M S_I (z) \sqrt{g} (R + \lambda^{-2}) .  \end{equation}  This is of course the Euclidean action for JT gravity, and we must choose $\lambda^2$ positive in order to put a constant curvature metric on the multi-mouthed Pacman geometries.
Note that $S_I (z) \equiv \sum_X S_I (X,z)  $ which is the only linear combination of the individual entropies that is invariant under any group (like discrete translations or rotations, or the full permutation group) that exchanges the regions $X$.  

Our derivation of two dimensional "gravitational" path integrals on multi-throat geometries from hydrodynamics does not in any obvious way lead to geometries of higher genus for a fixed number of boundaries.   From the Euclidean gravity point of view considerations of two dimensional locality are sometimes invoked to explain the necessity of including them.  The remarkable paper of\cite{sss} gives us a clear picture of what their role is.   We have explained connected contributions to multipoint spectral form factors in terms of time averaging and the hydrodynamic approximation.   Alternatively, they can arise from averaging over random choices for the boundary Hamiltonian.   In that case, higher genus contributions are just higher order terms in the $e^{-S}$ expansion of the random matrix problem.  In the two point spectral form factor, they contribute to the behavior in the plateau region, which kicks in at time scales of order $e^S$.   These are however time scales over which we expect the hydrodynamic approximation to break down, since it was based on neglecting energy differences of order $e^{-S}$.  The Markov transition kernel $W$ does not contain enough information to follow the system over the time scales on which higher genus contributions become important.   It can tell us about the ramp region of the SFF\cite{swingle} but not about the plateau.  Note however that hydrodynamics does tell us about the spectral density of the Hamiltonian, which means that it determines all the higher genus contributions to the random matrix problem. Thus hydrodynamic information, combined with the hypothesis that behavior on the plateau is well approximated by some random matrix theory, tells us what the random matrix theory is.  

In the context of the present class of models, it is clear that the relevant hydrodynamic and ramp time scales are determined by the entropy of single blocks of the system rather than the system as a whole, for which $e^S$ is unimaginably larger.   Thus our picture of thermalization and the onset of chaos in a model that lives on a graph, is that on a time scale of order $L^2$ in microscopic units, spectral correlation functions, averaged over the parametrically smaller time $\delta t$, are well approximated by the equations of fluctuating hydrodynamics.   This persists until times of order $e^{S_{block}}$, when the details of the finely grained block energy level spacings become important.  It is plausible that in any sort of realistic system, the behavior of spectral correlators is well approximated by random matrix theory, with a spectral density that already appears in the hydrodynamic equations.

\section{Towards models with Gravity duals}

Most of the models we have discussed, do not have {\it Einstein-Hilbert (EH) duals}.  That is, despite the generic appearance of a two dimensional gravitational Lagrangians, they do not correspond to models of quantum gravity whose observable correlation functions can be calculated approximately by solving gravitational field equations with a two derivative action.  Models in which there is only one block $X$ are an illustrative example.  These models have no hydrodynamic transport at all. In order to fall into the class of SYK models, whose dynamics is known to be well approximated by classical gravitational calculations, one must first of all insist that the boundary Hamiltonian $H(P_S , S)$ is negligible compared to the entropy $S$.   In addition, on a microscopic level invisible to hydrodynamics, the dynamics must have the property of {\it fast scrambling}\cite{hpss}.   The lattice field theory models that were the focus of\cite{tblucas} do not fall into this category.  It is not clear at this stage whether these two necessary conditions for the existence of an EH dual are related to each other.  If they are both satisfied, for a model with only a single block, then it appears that the model is of the SYK form and its coarse grained properties are described by Euclidean quantum gravity.

For models with a large number of blocks $X_i$, transport of entropy throughout the system is controlled by the "metric" 
 $$ G[e(X_i), e(X_j)] . $$  if that metric has the locality properties of a lattice in a compact  $n$ dimensional space (in the sense that the associated Fokker-Planck equation describes diffusion on that lattice) then we might be looking at a tensor network approximation to quantum gravity in $AdS_{n+2} \times {\cal K} $, where ${\cal K}$ is a compact manifold.  In order for that to be true, at least three criteria must hold
 \begin{itemize}
 
 \item  We must be able to imbed the model in a sequence of models with exponentially increasing numbers of degrees of freedom, with entangling tensors between successive models local on the lattice.  
 
 \item  The equations must be invariant under a discrete group of lattice symmetries, which is a subgroup of $SO(n + 1)$ .   The subgroup must grow with the number of degrees of freedom in the sequence of tensor network models.  This subgroup must be a dihedral group, the symmetry group of an anti-prism.   The formalism of\cite{tblucas} needs to be generalized to take into account the translational symmetries and the transport of discrete momentum around the lattice.  
 
 \item  This is the most obscure and difficult criterion of all:  angular momentum at a point on a horizon appears as linear momentum in the tangential directions to the holographic screen.  The momentum density satisfies a universal Navier-Stokes equation for an incompressible fluid, on all horizons.   For large AdS black holes there are also compression waves with wavelength longer than the AdS radius.   Incompressibility on sufficiently short scales is a direct consequence of fast scrambling, while the compression waves tell us about the field theory nature of the large scale dynamics in AdS space.  The puzzling question is how to derive a local diffusive equation for momentum density from a fast scrambling system.  Models of the SYK family do not have any obvious answer to this puzzle.
 
 \end{itemize}
 
 \section{Conclusions}
 
 We have shown how to write the time averaged connected spectral correlators
 \begin{equation} {\rm Tr }\ \rho_{TFD} (t_1) \ldots \rho_{TFD} (t_n) \end{equation} of quite general quantum many body systems
in terms of Euclidean functional integrals of gravitational and scalar fields on a two dimensional negatively curved surface with the topology of a disk with $n$ boundaries.  These connected correlators vanish at the microscopic level but are non-zero after time averaging.  Our analysis assumes that the hydrodynamic approximation\cite{tblucas} to the evolution of the diagonal matrix elements of the density matrix in the "block energy basis" is valid.  The authors of\cite{swingle} have shown that hydrodynamics reproduces the "ramp" in the two point SFF.   One consequence of our analysis is that we can expect the hydrodynamics, and these two dimensional gravity representations, to break down on time scales longer that $e^{S(X)}$ where $S(X)$ is the entropy of a single large block of the system.  However, the results of \cite{sss} indicate that if we make the assumption that after that point, time averaged quantities are approximately the same as those in any random system with the same spectral density, Euclidean path integrals of higher topology will compute all the terms in the $e^{-S}$ expansion of connected correlators. 

We briefly discussed the requirements for this formal representation of connected correlators to actually be an analog of a true AdS/CFT duality.  For systems with a single block $X$ it is essentially the requirement that the system be a member of one of the SYK ensembles, although there is no real sense in which these models have an EH dual\footnote{In a paper to be published soon\cite{qg2d}, Draper and the present author have constructed a model in $AdS_2$ that has an EH dual away from a small region near the horizon, but has eigenvalue statistics like those of an SYK model.}.   For systems with multiple blocks the problem of constructing a quantum model with an EH dual in $AdS_{d > 2}$ is more formidable.

\end{document}